\documentclass[a4paper]{article}
\usepackage{graphicx}
\usepackage{twocolceurws}
\usepackage{float}

\usepackage{hyperref}
\hypersetup{
    colorlinks=true,
    linkcolor=blue,
    filecolor=magenta,      
    urlcolor=magenta,
}

\title{Characterizing COVID-19 Misinformation Communities Using a Novel Twitter Dataset}

\author{
Shahan Ali Memon \\ samemon@cs.cmu.edu
\and
Kathleen M. Carley \\  kathleen.carley@cs.cmu.edu
}

\institution{Carnegie Mellon University}

\begin{document}
\maketitle
\begin{abstract}
From \emph{conspiracy theories} to \emph{fake cures} and \emph{fake treatments}, COVID-19 has become a hot-bed for the spread of misinformation online. It is more important than ever to identify methods to debunk and correct false information online. In this paper, we present a methodology and analyses to characterize the two competing COVID-19 misinformation communities online: (i) \emph{misinformed users} or users who are actively posting misinformation, and (ii) \emph{informed users} or users who are actively spreading true information, or calling out misinformation. The goals of this study are two-fold: (i) collecting a diverse set of annotated COVID-19 Twitter dataset that can be used by the research community to conduct meaningful analysis; and (ii) characterizing the two target communities in terms of their network structure, linguistic patterns, and their membership in other communities. Our analyses show that COVID-19 misinformed communities are denser, and more organized than informed communities, with a possibility of a high volume of the misinformation being part of disinformation campaigns. Our analyses also suggest that a large majority of misinformed users may be anti-vaxxers. Finally, our sociolinguistic analyses suggest that COVID-19 informed users tend to use more narratives than misinformed users.
\end{abstract}

\section{Introduction}
With the emergence of COVID-19 pandemic, the political and medical misinformation has elevated to create what is being commonly referred to as the \emph{global infodemic}. False information has hampered proper communication, and affected the decision-making process \cite{baines2020defining}. This makes debunking of false information vitally important. According to one study \cite{tan2015exposure}, if left undisputed, misinformation can in fact exacerbate the spread of the epidemic itself. Process of debunking misinformation, however, is complex and one that is not completely understood \cite{chan2017debunking}. This is because in order to conduct any intervention, it is first imperative to be able to identify the misinformation, as well as the misinformed communities. Because of the scarcity of data, and diversity of misinformation themes, this is already a challenging task in itself, but is also not enough. A second, and arguably a more important aspect of an intervention is to be able to \emph{correct} and \emph{change} the beliefs of the misinformed communities. To be able to do this, it is important to understand how different communities interact, which sub-communities they belong to, and what are their preferences. In this paper, we characterize the COVID-19 misinformation communities on Twitter in terms of their network structure, linguistic patterns, and membership in other misinformation and disinformation sub-communities. In the process, we also design and collect a large annotated dataset with a comprehensive codebook that we make available for the community to use for further analysis and models for misinformation detection.


\section{Background}
\subsection{COVID-19 Datasets}
In the short amount of time, many COVID-19 datasets have been released. Most of these datasets are generic, and lack annotations or labels. Examples include multilingual corpus on a wide variety of topics related to COVID-19 \cite{chen2020tracking,abdul2020mega,huang2020coronavirus}, longitudinal Twitter chatter dataset \cite{banda2020large}, multilingual dataset with location information of the users \cite{qazi2020geocov19}, Twitter dataset for Arabic tweets \cite{alqurashi2020large}, Twitter dataset for popular Arabic tweets \cite{haouari2020texttt}, and dataset for identification of stance, replies, and quotes \cite{villa2020stance}. Most of these datasets either have no annotations at all, employ automated annotations using transfer learning or semi-supervised methods, or are not specifically designed for misinformation. 

In terms of datasets collected for COVID-19 misinformation analysis and detection, examples include CoAID \cite{cui2020coaid} which contains automatic annotations for tweets, replies, and claims for fake news; ReCOVery \cite{zhou2020recovery} is a multimodal dataset annotated for tweets sharing reliable versus unreliable news, annotated via distant supervision; FakeCovid \cite{shahi2020fakecovid} is a multilingual cross-domain fake news detection dataset with  manual annotations; and \cite{dai2020ginger} is a large-scale Twitter dataset also focused on fake news. A survey of the different COVID-19 datasets can be found in \cite{latif2020leveraging} and \cite{shuja2020covid}.

In terms of the diversity of the classes, and the size of the dataset, the most relevant dataset is by Alam et al. \cite{alam2020fighting} who, like our study, present a comprehensive codebook to annotate tweets on a finer granularity. Their dataset, however, is limited to a few hundred tweets, and our dataset is much more diverse in the range of topics covered. Dharawat et al. \cite{dharawat2020drink} present a similar dataset with focus on the severity of the misinformation. However, their dataset does not consider the different ``types" of misinformation. Finally, Song et al. present a dataset in \cite{song2020classification} which contains a diverse set of 10 categories, but still is not as large, and contains fewer categories in relation to the dataset collected within our study.

\subsection{Misinformation Analysis}
A plethora of research has already been conducted for analysing COVID-19 misinformation online. Some examples include categorization and identification of misinformed users based on their home countries, social identities, and political affiliation \cite{huang2020disinformation,sharma2020covid}, characterization of different types conspiracy theories propagated by Twitter bots \cite{ferrara2020types}, characterization of the prevalence of low-credibility information related to COVID-19 \cite{yang2020prevalence}, exploratory analysis of the content of COVID-19 tweets \cite{ordun2020exploratory,shahi2020exploratory}, understanding the types, sources, and claims of COVID-19 misinformation \cite{brennen2020types}, and comparison of the credibility of COVID-19 tweets to datasets pertaining to other health issues \cite{broniatowski2020covid}. To the best of our knowledge none of the studies have characterized COVID-19 misinformation communities in terms of their sociolinguistic patterns. In this study, we do not characterize the misinformation \emph{content} directly. Instead, we conduct a set of analysis to understand and characterize the competing COVID-19 \emph{communities} through their content, and content-sharing behaviors and interactions. 

\section{Methodology}
\subsection{Data Collection}
To collect Twitter dataset, we use Twitter search API using a diverse set of keywords as shown in table \ref{tab:keywords-covid} to collect data. We collected our data on three days: 29th March 2020, 15th June 2020, and 24th June 2020. Each of these collections extracted a set of tweets from their corresponding week. For the annotation process, tweets were randomly sampled from that set.

\begin{table}[!htb]
\caption{This table shows the hashtags, and keywords we used in conjunction with ``coronavirus" and ``covid" to collect data from Twitter}
\resizebox{\columnwidth}{!}{
\begin{tabular}{|c|c|} 
\hline
Type & Terms \\
\hline
\hline
Keywords & \begin{minipage}{20em}\textit{bleach, vaccine, acetic acid, steroids, essential oil, saltwater, ethanol, children, kids, garlic, alcohol, chlorine, sesame oil, conspiracy, 5G, cure, colloidal silver, dryer, bioweapon, cocaine, hydroxychloroquine, chloroquine, gates, immune, poison, fake, treat, doctor, senna makki, senna tea}\end{minipage} \\
\hline
Hashtags & \begin{minipage}{20em}\textit{\#nCoV20199, \#CoronaOutbreak, \#CoronaVirus, \#CoronavirusCoverup, \#CoronavirusOutbreak, \#COVID19, \#Coronavirus, \#WuhanCoronavirus, \#coronaviris, \#Wuhan}\end{minipage}\\
\hline
\end{tabular}
}
\vspace{1em}
\label{tab:keywords-covid}
\end{table}
\vspace{-4mm}
\subsection{Data Annotation}
Our annotation task aims to determine the category to which a given tweet belongs to. After many discussions and revisions, we identify 17 categories that a particular tweet could classify to. These 17 categories are defined in table \ref{tab:coding-scheme-1}. These categories are defined in further detail along with their definitions and examples in our codebook which we make available for the public to use. 

\begin{table}[!ht]
\tiny
\caption{This table describes the categories we identified to classify/annotate tweets along with the distribution of annotations as identified by Annotator 1 in the first phase.}
\centering
\setlength\tabcolsep{10pt}
\resizebox{\columnwidth}{!}{\begin{tabular}{|c|c|} 
\hline
Category & Count\\
\hline
\hline
Irrelevant & 131\\
\hline
Conspiracy & 924\\
\hline
True Treatment & 0\\
\hline
True Prevention & 175\\
\hline
Fake Cure & 141\\
\hline
Fake Treatment & 34\\
\hline
False Fact or Prevention & 321\\
\hline
Correction/Calling out & 1331\\
\hline
Sarcasm/Satire & 476\\
\hline
True Public Health Response & 163\\
\hline
False Public Health Response & 3\\
\hline
Politics & 512\\
\hline
Ambiguous/Difficult to Classify & 143\\
\hline
Commercial Activity or Promotion & 37\\
\hline
Emergency Response & 17\\
\hline
News & 95\\
\hline
Panic Buying & 70\\
\hline
\end{tabular}}
\vspace{2em}
\label{tab:coding-scheme-1}
\end{table}

Based on these categories, tweets were randomly and uniformly sampled from the data collection to maintain diversity in terms of topics covered. In the first phase around 4573 tweets were annotated by a single annotator. Table \ref{tab:coding-scheme-1} shows the distribution of the data in terms of the different categories as annotated by the first annotator. In the second phase, 651 of these annotated tweets were assigned randomly to 6 other annotators.


\section{Data Description}
Our data collection strategy is different from others in two main aspects: (i) we have a diverse set of categories taking into consideration different types of information and misinformation online; and (ii) our dataset is one of the very few, if not the only one, with emphasis on informed communities with categories such as ``True Prevention", ``Calling out/correction", ``True Public Health Response", and ``Sarcasm". We believe this is necessary as building models requires not just the annotation of false information, but as well as complementary true information categories. 

\begin{figure}[!htp]
\caption{This chart shows the frequency of each identified topic across all the tweets. Note: Some tweets may have more than one topic.}
    \centering
    \includegraphics[width=8cm]{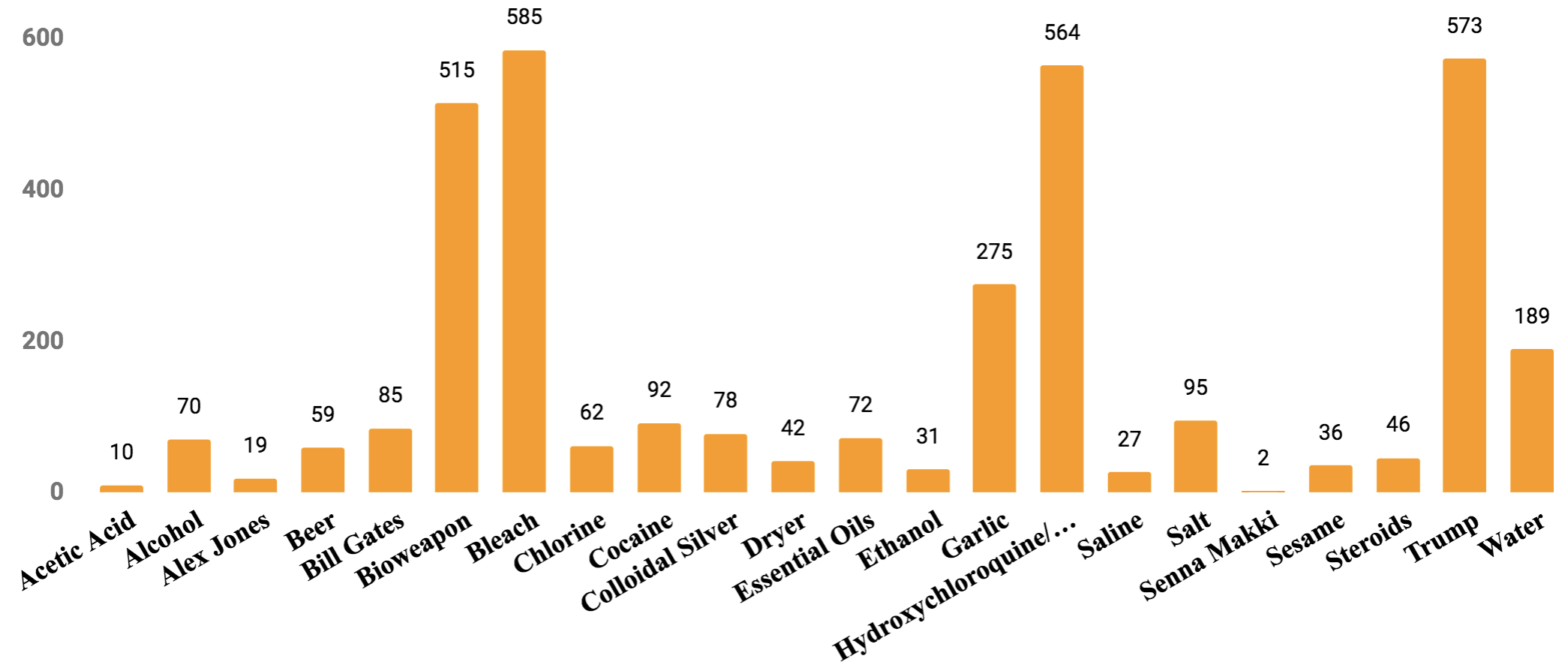}
    \label{fig:data-categories}
\end{figure}

At the end, we have $4573$ annotated tweets, comprising of $3629$ users with an average of $1.24$ tweets per user. Our annotated data not only covers a wide range of categories as observed in table \ref{tab:coding-scheme-1}, but also covers a wide range of topics as can be seen in figure \ref{fig:data-categories}. We call this dataset \emph{CMU-MisCOV19} \cite{cmumiscov19}. In adherence to the FAIR principles, the database and the codebook has been uploaded to Zenodo and is accessible with the following link: \href{http://doi.org/10.5281/zenodo.4024154}{http://doi.org/10.5281/zenodo.4024154}. In adherence to the Twitter's terms and conditions, we do not provide the full tweet JSONs, but provide the tweet IDs so that the tweets can be rehydrated. We also provide the annotations, and the date of creation for each tweet for the reproduction of the results of our analyses. The annotated tweets are included in a CSV file with the following fields: \textit{status\_id} (tweet id of the tweet), \textit{status\_created\_at} (timestamp of the creation of the tweet), \textit{annotation1} (annotated class of the tweet by the first annotator), and \textit{annotation2} (annotated class of the tweet by the second annotator, if exists).

\section{Analysis and Discussion}
\subsection{Identifying Communities}
Conducting analyses for a competing set of communities requires identifying those communities first. Because we have already annotated data across a set of true and false information categories, we identify the membership of the users by assigning a valence of +1 to the categories \emph{True Treatment, True Prevention, Correction/Calling Out, Sarcasm/Satire, and True Public Health Response}, and a valence of -1 to the categories \emph{Conspiracy, Fake Cure, Fake Treatment, False Fact or Prevention, and False Public Health Response}. Note that we assign the valence to the categories (or annotations) and not the tweets themselves. This is so that we can leverage the annotations from multiple annotators. At the end, we compute the valence of each user as a weighted sum of the valence of the annotations assigned to their tweets. Then we use the valence assigned to each user to identify their membership i.e. if valence is greater than 0, the user is assigned to the \emph{informed} group, and if the valence is less than 0, the user is assigned to the \emph{misinformed} group. Out of 3629 users, the community detection process assigns 47\% (1697) of the users to the informed group, 29\% (1043) of the users to the misinformed group, and 24\% (889) of the users to ambiguous or irrelevant category\footnote{Irrelevant users are users who have only posted tweets within other categories such as ``Politics" or ``Emergency". Because these categories do not have an assigned valence related to misinformation, they are not relevant for the purposes of this study.}. 

\subsection{Data Augmentation}
Because our goal is to characterize communities and their behaviors, once we identify the two communities, we collect the timelines of users in each community to augment our data. Our hypothesis is that these additional posts can be used to mitigate survivorship bias \cite{brown1992survivorship} within our analyses. To conduct network analysis, bot analysis, and sociolinguistic analysis, we first extract only the COVID-19 related tweets from the timelines of each user. We do this by filtering all the tweets by the case-insensitive keywords \emph{``corona"} and \emph{``covid"}. This yields a total of 330609 tweets with an average of 91 tweets per user.

\subsection{Network Analysis}
\begin{figure}[!htb]
\caption{Retweet+Mention+Reply network with informed users (in green) and misinformed users (in red) created using ORA-PRO \cite{altman2018ora,carley2017ora}. Note: Users with unidentified or ambiguous membership have been removed from the graph for simplicity.}
    \centering
    \includegraphics[width=4cm]{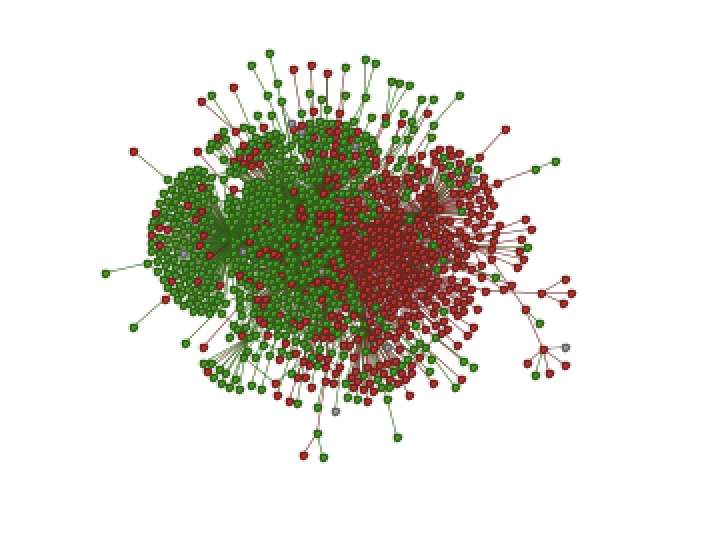}
    \label{fig:covid-network}
\end{figure}

To conduct network analysis, we extract the retweet, mention, and reply networks of the two target communities, and combine those networks together. We then compute the \emph{network density} for each of the two groups. As described in \cite{memon2020characterizing}, network density is defined as the ratio of actual connections and potential connections. In dense networks, conformity of the ideas is highly encouraged, and difference of opinions is discouraged. We also use ORA-PRO \cite{carleyora,altman2017ora,altman2018ora} to plot the network graph as shown in figure \ref{fig:covid-network}

We note that both the informed and misinformed users display echo-chamberness with misinformed sub-communities being much denser than the informed sub-communities as shown in table \ref{tab:density-covid}. We do, however, notice some two-way communication from both sides. 

\begin{table}[!ht]
\small
\caption{This table shows the number of nodes, links, and the network density for the two target sub-communities.}
\centering

\begin{tabular}{|c||c|c|c|} 
\hline
Measure & Overall & Informed & Misinformed\\
\hline
\hline
Nodes & 2477 & 1515 & 923\\ 
Links & 2947 & 1489 & 826\\ 
Network Density & 4.8e-4 & 6.5e-4 & 9.7e-4\\
\hline
\end{tabular}
\vspace{1em}
\label{tab:density-covid}
\end{table}

\begin{figure}[!htp]
\caption{Retweet (left), mention (middle), and reply network (right) with informed users (in green) and misinformed users (in red) created using ORA-PRO. \cite{altman2018ora,carley2017ora}}
    \centering
    \includegraphics[width=2.5cm]{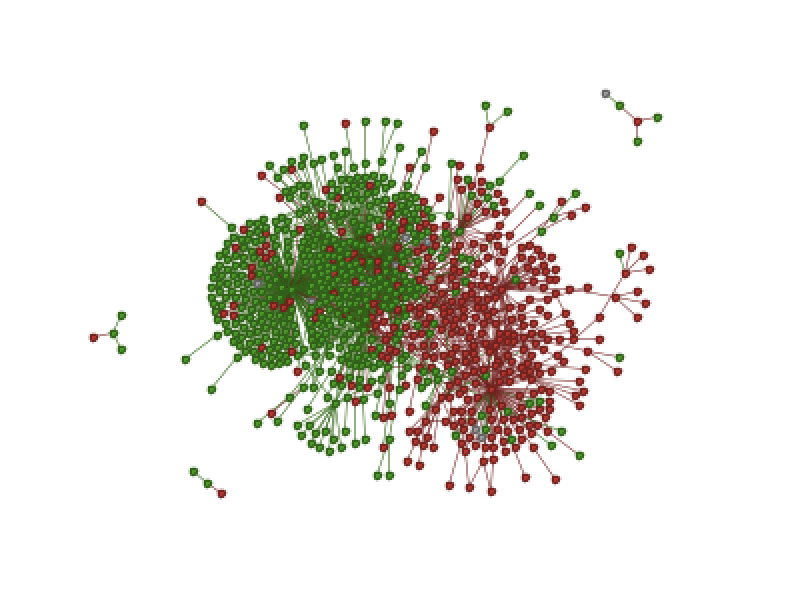}
    \includegraphics[width=2.5cm]{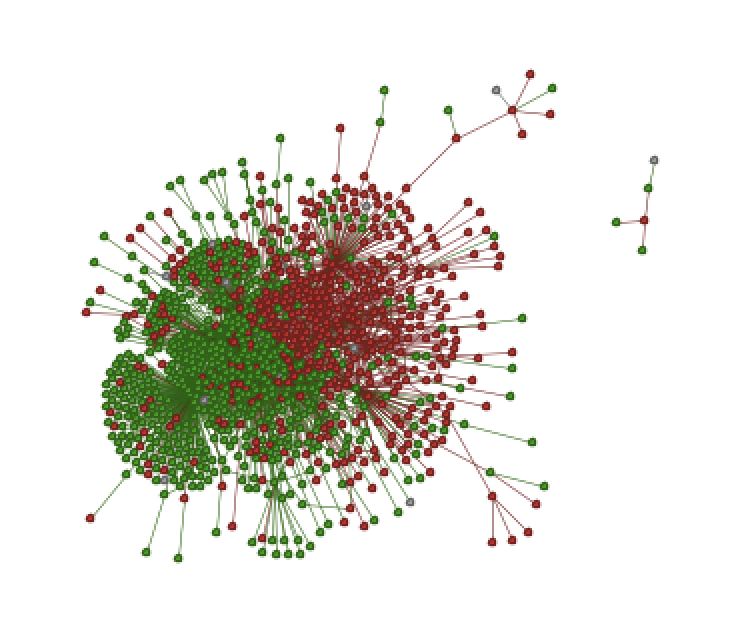}
    \includegraphics[width=2.5cm]{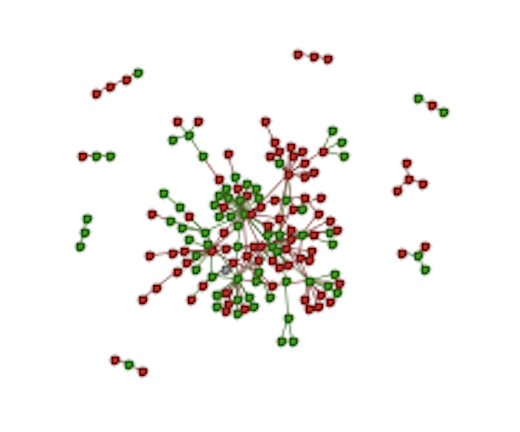}
    \label{fig:covid-network-separate}
\end{figure}

We also plot the retweet, mention and reply network separately as shown in figure \ref{fig:covid-network-separate}. While retweet, and mention network show little to no two-way communication, we can observe that the reply network, while small in size, does in fact have much more inter-group engagement. We hypothesize that this is likely a consequence of the ``corrective" or ``calling-out" behavior.

\subsection{Bot Detection}
To understand the role of bots within the two competing groups, we used Bot-Hunter \cite{beskow2018bot,beskow2018bot2,beskow2018introducing,beskowspringer}, which has a precision of .957 and a recall of .704, to identify potential bot-like accounts. We use the probability of greater than or equal to .75 as our confidence threshold to identify bots. We use a two-sample z-test for the difference of proportions ($\alpha = 0.05$) to test the difference in proportion of bots between the two competing groups of users. The results of our analyses can be found in table \ref{tab:botanalysis}. 

\begin{table}[!ht]
\small
\caption{This table shows the number and percentage of bots within each of the two competing groups}
\centering

\begin{tabular}{|c||c|c|c|} 
\hline
Measure & Overall & Informed & Misinformed\\
\hline
\hline
Number of Users & 3629 & 1697  & 1043\\ 
Number of Bots & 505 & 184 & 202\\ 
Percentage of Bots & 14\% & 11\% & 19\%\\
\hline
\end{tabular}
\vspace{1em}
\label{tab:botanalysis}
\end{table}

We observe that from a total of 3629 users, 14\% (505) of the users are identified as bots. The percentage of bots within identified misinformed users, however, is much higher (19\%) than within identified informed users (11\%). We find our results to be statistically significant ($p<0.001$; $z=-6.23$). This indicates that more than 1/5th of the misinformation related posts in our dataset are potentially a result of disinformation campaigns related to COVID-19. 

\subsection{Sociolinguistic Analysis}

\begin{table*}[!ht]
\caption{This table shows the summary of our analyses across all the linguistic dimensions described above using LIWC. The first column shows the lexical category. The second and third columns show the test statistic ($M_1$) as the mean of the LIWC indices for informed and misinformed communities respectively. The fourth and fifth columns display the z-score and p-value for the independent z-test for the difference in means.}
\centering
\begin{tabular}{|c||c|c|c|c|} 
\hline
Lexical Category & $M_1$ (Informed) & $M_1$ (Misinformed) & z-score ($Z_1$) & p-value ($Z_1$) \\
\hline
\hline
function & 33.90 & 29.32 & 7.25 & $<.001$\\
\hline
\hline
pronoun & 7.97 & 6.53 & 4.89 & $<.001$\\
\hline
ipron & 3.26 & 3.03 & 1.23 & $.2$\\
\hline
ppron & 4.71 & 3.49 & 5.39 & $<.001$\\
\hline
Analytic & 69.83 & 76.01 & -4.82 & $<.001$\\
\hline
social & 6.49 & 5.05 & 5.45 & $<.001$\\
\hline
family & .34 & .20 & 2.24 & $.03$\\
\hline
friend & .17 & .17 & -.03 & $.97$\\
\hline
Authentic & 25.12 & 16.43 & 6.78 & $<.001$\\
\hline
Tone & 35.42 & 37.59 & -1.45 & $.15$\\
\hline
informal & 4.89 & 5.16 & -1.63 & $.10$\\
\hline
swear & .51 & .34 & 1.86 & $.06$\\
\hline
\hline
\end{tabular}
\vspace{1em}
\label{tab:liwc-covid}
\end{table*}

To understand the linguistic differences between the two competing communities, we conduct a linguistic analysis based on the tweets of the two groups by using the Linguistic Inquiry and Word Count (LIWC) program \cite{pennebaker2015development}. LIWC is a text analysis tool which looks at the different lexical categories each of which is psychologically meaningful. For a given text, LIWC calculates the percentage of each LIWC categories. All of these categories are based on word counts. 

We run the LIWC program on the timelines of all the members for each of the two competing groups. We only use tweets relevant to COVID-19. We also remove users identified as bots. Because some users may be more active than others, using the results of the program as is may introduce biases in our analyses. To account for those biases, we first normalize the percentages by the size of the data for each user. We use the mean of the normalized LIWC indices of tweets of individual users for a given lexical category as our test statistic. We use an independent z-test for the difference in means to establish statistical significance. For all our tests, $\alpha=0.05$. Our analyses are summarized in table \ref{tab:liwc-covid}.

For this part, we focus on investigating three linguistic dimensions, each of which, along with its linguistic correlates, is described below.

\subsubsection{Narrative Discourse Structure}
Narratives play a central role in how individuals process information, communicate, and reason \cite{veselkova2017narrative}. We set to test the differences in the usage of narratives or anecdotes between the two COVID-19 misinformation communities. The LIWC correlates for narrative discourse structure include high usage of function words, pronouns, analytic summary dimension, and authenticity. High usage of function words and pronouns happens more often when expressing feelings and behaviors which tends to happen frequently in narratives \cite{pennebaker2011secret}. Moreover, low analytical thinking also suggests narrative language \cite{pennebaker2015development}. Furthermore, authentic individuals tend to be more personal, humble, and vulnerable \cite{pennebaker2015development}. Therefore, we use all of these as proxies to identify variation in the use of narratives across communities. 

In the past \cite{memon2020characterizing}, it has also been suggested that misinformed communities (eg. anti-vaxxers) tend to use many more pronouns suggesting highly narrative discourse structure. In this analysis, however, we find that informed users in the COVID-19 discourse use significantly more pronouns, more functional words, mention more family-related keywords, are less analytical, and more authentic and honest in comparison to misinformed users. All of these suggest that informed users may use many more narratives than misinformed users. This is an interesting finding as it presents a dichotomy between the different misinformation communities (eg. anti-vaxxers and COVID-19 misinformed community). In hindsight, this is also an intuitive result, as our informed group is obtained from corrective discourse where users present their stories of family members or friends suffering from COVID-19 to call out conspiracies and false information. Because the two communities still seem to have less two-way communication, this also suggests that just the content and framing of the message (i.e. narratives) may not be enough, and perhaps there is a need to connect the two groups by identifying an effective medium of communication.

\subsubsection{Tone}
Tone describes how positive a given text is. According to the definition by LIWC, the higher the tone index, the more positive the tone. Indices less than 50 typically suggest a more negative tone. While we do not see significant differences in the emotional tone of the competing groups, we find both the communities to be highly negative.

\subsubsection{Linguistic formality} Formality of the language has often been considered as one of the most important dimensions for stylistic variation. In \cite{graesser2014coh}, authors define linguistic formality as a style of writing that is meant to be precise, coherent, articulate and convincing to an educated audience, as opposed to informal discourse which is filled with deictic references (eg. here, there), pronouns, and narration. The LIWC correlates to this dimension are swear words (swear), and informal language (informal). Informal language in LIWC is computed on the bases of swear words, netspeak (eg. btw, lol), nonfluencies (eg. err, hmm), assents (eg. agree, OK), and fillers (eg. youknow). 

From table \ref{tab:liwc-covid}, it can be observed that misinformed users tend to be more informal than informed users, though informed users tend to use more swear words than misinformed users. This is intuitive as many of our informed users post corrective or sarcastic tweets to call out misinformation. However, our results are not significant, and, hence inconclusive. 


\subsection{Vaccination Stance}
To understand the interplay between the different kinds of misinformation themes and communities, we identify the vaccination-related stance of the members of the misinformed sub-community. To do that, we first identify the subset of misinformed community who have posted at least one tweet related to ``vaccines" in the past. We then collect the user-to-hashtag co-occurrence network. We use the valence of the vaccination hashtags obtained via the label propagation-based method mentioned in the study in \cite{memon2020characterizing} to identify the stance of each member (pro vs. anti) based on the weighted sum of the valences of the hashtags. If the weighted sum is greater than 0, we identify the member as pro-vaxxer, and if the weighted sum is less than 0, we identify the member as anti-vaxxer. The distribution of the pro- and anti-vaxxers within the COVID-19 misinformed group is as shown in table \ref{tab:vacc-covid}.

\begin{table}[!ht]
\small
\caption{This table shows the number and percentage of pro- and anti-vaxxers within the misinformed group.}
\centering

\begin{tabular}{|c|c|} 
\hline
Measure & Value\\
\hline
\hline
Users w/ vaccine-related tweets & 2750 (out of 3629)\\ 
Misinformed users & 1027 (37\%)\\ 
Anti-vaxxers & 423 (41\%)\\
Pro-vaxxers & 224 (22\%)\\
Ambiguous & 380 (37\%)\\
\hline
\hline
Misinformed pro-vaxxer bots & 37 (17\%)\\
Misinformed anti-vaxxer bots & 82 (22\%)\\
\hline
\end{tabular}
\vspace{1em}
\label{tab:vacc-covid}
\end{table}

We observe that from 1027 COVID-19 misinformed users in our dataset, 41\% of the members are identified as anti-vaxxers, whereas only 22\% of the members  are identified as pro-vaxxers. The difference between the proportions of the two communities is significantly high. We also identify the proportion of bots within each of the two groups: \emph{misinformed pro-vaxxers}, and \emph{misinformed anti-vaxxers}. As shown in table \ref{tab:vacc-covid}, 17\% of the misinformed pro-vaxxers are bots, which is significantly lower than the proportion of bots within the misinformed anti-vaxxers. The first thing this suggests is that a big chunk of COVID-19 misinformation online may in fact be \emph{disinformation}, and hence, intentional. The existence of bots within both the informed and misinformed communities also suggests that much of the disinformation online may be an organized effort to amplify the COVID-19 debate to create discord in the communities as seen in the past with Twitter bots and Russian trolls \cite{broniatowski2018weaponized}. 

\section{Limitations}
The first important limitation pertaining to our work is that most of our analyses are based on the data that has been annotated by only 1 annotator. We try to mitigate this by having more than 1/7th of our annotations annotated by a second annotator, and taking into account all those annotations while computing the membership for each user. Another limitation to our work is that all our analyses are correlational in nature, and do not depict causation. A limitation pertaining to  our data collection strategy is that we collect our data across a period of three weeks, augment our data with timelines of users, and update our list of hashtags to account for new themes. We then sample a subset of this data for annotation process. Because of the way data was collected, it cannot be used for assessing change over time. Moreover, while this ensures the diversity of misinformation-related topics and agents, it may limit our ability to estimate the actual extent to which the different types of stories are more or less present. Another limitation related to our bot analysis is that we use a second-level inference from a trained model. We try to mitigate this by using labels with probability greater than or equal to .75 to ensure high quality labels. Finally, unlike ``vaccination" related discourse, COVID-19 does not have a clear definition of the ``stance" of the users. This is because there are many sub-topics associated to COVID-19 each of which could have its own stance. In this work, we categorize users based on misinformation. However, the relationship between misinformation and stance vis-a-vis issues is complex, and one that needs to be understood. In the future work, we hope to explore this relationship to create a systematic way of characterizing communities both in terms of misinformation, and the difference stances of the users. 

\section{Conclusion}
In this paper, we present a methodology to characterize the competing COVID-19 misinformation communities by comparing them in terms of their network structure, sociolinguistic variation, and membership in disinformation campaigns and in other health-related misinformation communities such as anti-vaxxers. We find that even though COVID-19 is a recent event, misinformation related to it has created a set of polarized communities with high echo-chamberness. Misinformed communities are observed to be denser than informed communities which is in line with previous studies such as \cite{memon2020characterizing}. We find that bots exist in both the informed and misinformed groups, but the percentage of bots in misinformed users is significantly higher suggesting the prevalence of disinformation campaigns. Our sociolinguistic analysis suggests that both the target communities depict negative emotional tone in their posts, with signals that informed users use many more narratives than misinformed users. Finally, we discover that many misinformed users may be anti-vaxxers. Our analyses suggest that misinformation communities are much more complex as they are highly organized, and tend to be highly analytical. Unlike previous suggestions \cite{sangalang2019potential}, they may not be responsive to narrative correctives, and hence, a ``one size fits all" generic messaging intervention for debunking misinformation may not be a feasible solution. A successful intervention may require to identify, and ban the disinformation campaigns. It may also be useful to identify the right medium of communication to connect the two groups. This can be achieved by identifying users in misinformed communities who are not \emph{rebroadcasting}, or have high betweenness centrality to be messengers for disseminating factual information. It may also be useful to further understand the linguistic patterns and preferences of these communities to create an effective \emph{content} and \emph{framing} of the messaging.

\subsubsection{Acknowledgements}

This work was partially supported by a fellowship from Carnegie Mellon University’s Center for Machine Learning and Health to Shahan A. Memon. We thank David Beskow for access to his Bot-Hunter model for bot analysis. We also thank members of CMU's Center for Computational Analysis of Social and Organizational Systems (CASOS) for insightful comments and discussions related to the data codebook and its revisions.

\bibliographystyle{alpha}
\bibliography{cikm}

\end{document}